\pdfoutput=1

\documentclass[11pt]{article}

\usepackage[utf8]{inputenc}
\usepackage{graphicx}
\usepackage{booktabs}
\usepackage{array}
\usepackage{float}
\usepackage{placeins}

\usepackage{ACL2023}

\usepackage{times}
\usepackage{latexsym}

\usepackage[T1]{fontenc}

\usepackage[utf8]{inputenc}

\usepackage{microtype}

\usepackage{inconsolata}

\title{A Literature Review of Keyword Spotting Technologies for Urdu}

\begin{document}

\author{Syed Muhammad Aqdas Rizvi \\
  \texttt{25100166@lums.edu.pk}}
\maketitle

\begin{abstract}
    This literature review surveys the advancements of keyword spotting (KWS) technologies, specifically focusing on Urdu, Pakistan's low-resource language (LRL), which has complex phonetics. Despite the global strides in speech technology, Urdu presents unique challenges requiring more tailored solutions. The review traces the evolution from foundational Gaussian Mixture Models to sophisticated neural architectures like deep neural networks and transformers, highlighting significant milestones such as integrating multi-task learning and self-supervised approaches that leverage unlabeled data. It examines emerging technologies' role in enhancing KWS systems' performance within multilingual and resource-constrained settings, emphasizing the need for innovations that cater to languages like Urdu. Thus, this review underscores the need for context-specific research addressing the inherent complexities of Urdu and similar URLs and the means of regions communicating through such languages for a more inclusive approach to speech technology.
\end{abstract}

\section{Introduction}
Progress in the development of speech technologies has included the advancement of KWS; as the sophistication of KWS technologies advances, the need to adapt and innovate for multilingual contexts becomes crucial, especially for LRLs such as Urdu. This literature review aims to offer an exploration of the evolution of emerging techniques to highlight the significant impact of cutting-edge technologies; simultaneously, this review underscores the need for tailored solutions that can enhance speech technologies across diverse linguistic landscapes with resource constraints and phonetic richness, such as for that of Pakistan and Urdu.

\begin{table*}[ht!]
    \centering
    \caption{Some notable consolidated metrics for KWS technologies.}
    \begin{tabular}{>{\raggedright}p{0.25\linewidth} >{\raggedright\arraybackslash}p{0.25\linewidth} >{\raggedright\arraybackslash}p{0.4\linewidth}}
        \toprule
        \textbf{Source} & \textbf{Metrics} & \textbf{Explanation} \\
        \midrule
        EdgeCRNN - an edge-computing oriented model of acoustic feature enhancement for keyword spotting \citep{Wei_Gong_Yang_Ye_Wen_2021} & \begin{minipage}[t]{\linewidth}Accuracy: 97.89\% \par Model parameters: 0.59M\end{minipage} & \begin{minipage}[t]{\linewidth}Accuracy measures the proportion of correct predictions. \par Model parameters indicate the complexity and size of the model, with 0.59M representing the total number of trainable parameters in the model.\end{minipage} \\
        \midrule
        HEiMDaL - Highly Efficient Method for Detection and Localization of Wake-Words \citep{10097018} & FRR: 0.45\% @ 12 FA/hr & False Reject Rate (FRR) is the rate at which true keywords are incorrectly rejected. False Accept Rate (FA/hr) is the rate at which the system incorrectly identifies non-keywords as keywords per hour. \\
        \midrule
        Self-Supervised Speech Representation Learning for Keyword-Spotting with Light-Weight Transformers \citep{10095929} & Accuracy: 95.8\% on Google speech commands v2 dataset & \\
        \midrule
        Unified Keyword Spotting and Audio Tagging on Mobile Devices with Transformers \citep{10095534} & \begin{minipage}[t]{\linewidth}Accuracy: 97.76\% \par mAP: 34.09\end{minipage} & mAP (mean Average Precision) quantifies the precision across different recall levels. \\
        \midrule
        Exploring Transfer Learning for Urdu Speech Synthesis \citep{jamal-etal-2022-exploring} & \begin{minipage}[t]{\linewidth}MOS naturalness: 3.40 \par MOS intelligibility: 3.30\end{minipage} & MOS (Mean Opinion Score) naturalness and intelligibility provide subjective ratings from listeners on how natural and understandable the synthesized speech sounds. \\
        \midrule
        Massively Multilingual Speech (MMS) Project \citep{pratap2023scaling} & WER: 18.7\% on FLEURS-54, 20.1\% for Urdu & WER (Word Error Rate) is calculated as the ratio of the total number of errors (substitutions, insertions, deletions) made by the system to the number of words in the reference text, giving a measure of how accurately the system can recognize spoken words. \\
        \midrule
        Urdu Keyword Spotting System using HMM \citep{irtza2014urdu} & Accuracy: 94.59\% & \\
        \midrule
        An Unsupervised Spoken Term Detection System for Urdu \citep{iqbalunsupervised} & \begin{minipage}[t]{\linewidth}Precision: 91.50\% (cross-speaker) \par 79.20\% (same-speaker)\end{minipage} & Precision measures the ability of the keyword spotting system in identifying keywords correctly without false positives, shown here for both cross-speaker and same-speaker settings. \\
        \bottomrule
    \end{tabular}
    \label{tab:metrics_kws}
\end{table*}

\section{Evolution and Current Trends in Keyword Spotting Technologies}
KWS relied heavily on Gaussian Mixture Models (GMMs), whose performance was the benchmark in speech recognition technologies. This approach used a mixture of Gaussian distributions to model the variation of speech sounds. However, by 2012, Deep Neural Networks (DNNs) began replacing GMMs, providing a more versatile means of capturing complex patterns in audio signals through deep, layered representations; in particular, the adoption of Recurrent Neural Networks (RNNs), especially those utilizing Long Short-term Memory (LSTM) cells, further enhanced KWS by maintaining contextual information over longer sequences of speech, significantly improving recognition accuracy \cite{Beaufays_2015}. On the other hand, Query-by-Example (QbyE) search techniques\footnote{Identifying audio documents matching a spoken query exactly or with some variation.} have also been an avenue worth exploring. Systems utilizing Subsequence Dynamic Time Warping algorithms used dynamic programming techniques to handle spoken queries across audio documents without requiring external information \cite{article}.

Notwithstanding the need for efficient and effective models, a more recent approach was the development of models like EdgeCRNN—a model designed for real-time keyword spotting on edge devices. EdgeCRNN combined convolutional neural networks (CNNs) with recurrent layers, optimizing the balance between computational efficiency and predictive accuracy, particularly in noisy environments \cite{Wei_Gong_Yang_Ye_Wen_2021}. Furthermore, work by \citet{yang2022personalized} employed multi-task learning for KWS with speaker recognition based on realistic usage of KWS systems; one can expect a specific group of users using a specific device; thus, there is reason to develop systems that are not user-agnostic. Their work demonstrated that such an approach yields better results in KWS, which is a significant indication in the context of languages like Urdu, which are phonetically rich. For such languages, where there is room for variations in speech for the same keyword, multi-task approaches such as this may be very successful.
Moreover, the integration of end-to-end learning models where the entire model is trained to map from audio waveforms to target keywords directly has also appeared in the landscape. This approach simplified the traditional pipeline, which involved separate feature extraction and classification stages. Models like HEiMDaL focused on detection and capturing the precise localization of keywords within audio streams, utilizing hybrid CNN architectures that were both computationally efficient and effective in real-world scenarios \cite{10097018}.

Most recently, advancements in KWS have also been driven by the rise of self-supervised learning and transformer models. Methodologies like Self-Supervised Speech Representation Learning (S3RL) use unlabeled data to pre-train models that can be fine-tuned for specific tasks such as KWS. These methods significantly reduce the dependency on large labeled datasets, reducing the investment of time and money required, which is especially promising for LRLs. For example, lightweight transformer models were pre-trained using Auto-Regressive Predictive Coding (APC; learning speech representations via using past frames to predict future frames); these models have been shown to improve the accuracy and efficiency of KWS systems on resource-constrained devices \cite{10095929}. Furthermore, advances in unified models that handle keyword spotting and audio tagging on mobile devices using transformer architectures demonstrate the ongoing integration of sophisticated AI techniques into everyday technology. Such models offer a multi-functional approach that accommodates real-world applications like intelligent assistants more holistically and may be better suited for LRLs due to their multi-task approach \cite{10095534}. Another novel approach uses vision-inspired keyword spotting; the application of models that use dynamic module skipping in streaming conformer encoders, such as those described by \citet{10447485}, demonstrates enhancements in keyword spotting by reducing computational demands, thus maintaining efficient performance.

However, despite the global application of advanced keyword-spotting technologies, it is increasingly evident that the benefits of these innovations are not uniformly distributed across all languages, distinctly so in the context of LRLs.

\section{Advancements in Multilingual Keyword Spotting Technologies}

Research on LRLs presents unique and language-specific challenges and solutions for integrating KWS in a multilingual framework. \citet{Patel_DN_Fathima_Shah_C_Kumar_Iyengar_2018} employed both Gaussian Mixture Model-Hidden Markov Model (GMM-HMM) and Deep Neural Network-Hidden Markov Model (DNN-HMM) architectures in a study on Automatic Speech Recognition (ASR) and KWS in the context of Manipuri, a low-resource Indian language. This study is significant as it highlights the effectiveness of advanced modeling techniques in improving keyword detection in an LRL.

Adapting recent developments in speech technologies, a noteworthy advancement in multilingual KWS for LRLs has been the introduction of cross-lingual speech representation learning, mainly through models like XLS-R \cite{babu2021xlsr}. Via leveraging vast amounts of unlabeled data across numerous languages to learn universal speech representations, XLS-R has demonstrated its effectiveness by achieving remarkable performance in speech recognition tasks across various languages without extensive labeled data in each language. This approach is well-suited for LRLs, as it allows for using pre-trained models that are fine-tuned on smaller, language-specific datasets, significantly reducing the barrier to entry for developing robust KWS systems in less-represented languages.

In a similar vein, we also have the angle of transfer learning (a machine learning methodology in which knowledge learned from a task is re-used on a similar task), which has been effectively employed to improve KWS systems in LRLs. A recent study on Urdu speech synthesis used transfer learning to adapt models trained on high-resource languages to Urdu, demonstrating satisfactory results despite the low availability of speech data \cite{jamal-etal-2022-exploring}. This study shows that this methodology can be particularly beneficial for languages like Urdu, where collecting and annotating large datasets is challenging. Therefore, by utilizing pre-trained models from related tasks or languages, researchers can accelerate the development of KWS systems for LRLs with reduced resource expenditure and time consumption.

As one of the most significant developments in speech technologies, the Massively Multilingual Speech (MMS) project used self-supervised learning to bring modern speech technology to otherwise less-represented languages. It did so by pre-training its models on over 1,000 languages, producing the \texttt{wave2vec 2.0} embeddings, and bringing modern technology for speech synthesis, ASR, and language identification to many more languages than previously, including LRLs. Notably, the many linguistic differences that catering to several languages accompanies did not hinder results, which exceeded prior work by producing Word Error Rates (WER) as low as 18.7\% on the \texttt{FLEURS-54} benchmark, which is less than half of the previous cutting-edge. However, more work is still needed in improving performance in LRLs such as Urdu, improvement in which was not as significant with a 20.1\% WER, marking a mere \textasciitilde{11\%} reduction from the cutting-edge \citep{pratap2023scaling}. This can be attributed to the lack of Urdu speech data and specific attention to Urdu's rich phonetics and phonology.

Indeed, integrating KWS systems into multilingual contexts faces numerous challenges, including handling diverse linguistic features and the need for training data for many languages. This is especially true in regions with high linguistic diversity, like Pakistan. For such regions, developing multilingual models that can operate across multiple languages without significant loss in performance is an avenue to explore for these models to benefit from the shared learning of linguistic features across languages.

In light of these developments, it becomes evident that individual languages' unique characteristics and challenges require more focused studies. This is needed to tailor these broad approaches effectively, and Urdu is no exception.

\section{Progress and Challenges in Urdu Keyword Spotting}
Initial efforts in Urdu KWS have utilized traditional approaches such as HMMs. One study, in particular, developed an Urdu KWS system based on filler models (a component designed to detect and handle non-words or irrelevant sounds that are not of interest but may appear in the input audio) and a phoneme recognizer using the all-phone model ASR. This system integrated string matching algorithms to refine keyword detection, achieving a high overall system accuracy of 94.59\% \cite{irtza2014urdu}. This work highlights the potential of using domain-specific models and refined training datasets to increase the effectiveness of KWS systems in Urdu.

Due to Urdu's scarcity of transcribed speech data, researchers have recently explored unsupervised methods to advance KWS technologies. Remarkably, a study presented an unsupervised KWS system that leverages Segmental Dynamic Time Warping and GMMs to identify keywords in speech without needing labeled data. The system used the Phonetically Rich Urdu Speech (PRUS) corpus for training and demonstrated precision rates up to 91.50\% for cross-speaker scenarios and 79.20\% for same-speaker scenarios \cite{iqbalunsupervised}. This study attests that unsupervised learning can extract meaningful features from raw speech data of LRLs like Urdu, thus making it a viable approach for such languages.

Despite these advancements, the development of KWS systems for Urdu is still troubled with challenges, namely, (1) the lack of large, annotated datasets, hindering the application of more complex machine learning models that require extensive training data, and (2) the phonetic diversity of Urdu and its script complexity, posing unique challenges in speech processing.

\section{Future Works}
Based on the current development of speech technologies overall, it is clear that the future of tasks such as KWS is best powered by unsupervised setups involving transformers due to the richness of their representations of speech. Such architectures give promising results while mitigating the need to do as much pre-processing for input data. This is especially significant in LRLs like Urdu, which lack as much data as possible. Additionally, contingent on the developments of more lightweight systems that perform similarly well, deployment of such technology in areas reliant on such LRLs for communication is also simplified, allowing for more feasible options for on-device and especially client-server deployments. However, the caveat is the need for a large amount of data, as representation learning through transformer architectures is inherently data-intensive.

Given these circumstances, future work should prioritize procuring more data for LRLs. With sufficient data, these languages can benefit more from cutting-edge speech technologies. Doing so can allow for a more refined focus on phonetically rich languages, such as Urdu, when developing models and methodologies in speech technologies. Furthermore, a specific focus on particular tasks like KWS is also a promising direction to see where methodologies such as multi-task learning may be effective regarding specific languages. Lastly, another critical direction is through networks and systems by evaluating modern methods' scale and deployment feasibility. This may require insight into a given area's technological infrastructure and communicating in particular languages to create more tailored and inclusive deployments. Overall, there is excellent potential for multi-faceted research work for KWS and speech technologies for LRLs, like Urdu, and better-resourced languages.

\section{Conclusion}

This literature review has discussed technological advancements in KWS, specifically regarding Urdu. As seen through various studies and developments, KWS has transitioned from traditional models like Gaussian Mixture Models and Hidden Markov Models to more advanced techniques involving deep neural networks, multi-task learning, and unsupervised learning.

Nevertheless, developing effective KWS systems in multilingual contexts and for LRLs like Urdu remains challenging. Advancements in cross-lingual speech representation and transfer learning (including fine-tuning) hallmark crucial steps to making modern and complex KWS technologies accessible and effective across such landscapes. Even so, the persistent scarcity of large annotated datasets and the inherent linguistic complexities of LRLs such as Urdu warrants continued innovation and tailored research efforts.

In a holistic sense, this review reflects on the broader implications for inclusivity in speech technologies. Ensuring the fair and equitable advancement of KWS systems across diverse landscapes enhances accessibility and enriches the interaction between humans and technology. Current advancements give hope that, with continued research, it is possible to develop adaptable and resource-efficient models that can handle the heterogeneity of the global population despite challenges faced by languages such as Urdu.

\bibliographystyle{acl_natbib}
\bibliography{custom}

\begin{thebibliography}{14}
\expandafter\ifx\csname natexlab\endcsname\relax\def\natexlab#1{#1}\fi

\bibitem[{Babu et~al.(2021)Babu, Wang, Tjandra, Lakhotia, Xu, Goyal, Singh, von Platen, Saraf, Pino, Baevski, Conneau, and Auli}]{babu2021xlsr}
Arun Babu, Changhan Wang, Andros Tjandra, Kushal Lakhotia, Qiantong Xu, Naman Goyal, Kritika Singh, Patrick von Platen, Yatharth Saraf, Juan Pino, Alexei Baevski, Alexis Conneau, and Michael Auli. 2021.
\newblock \href {http://arxiv.org/abs/2111.09296} {Xls-r: Self-supervised cross-lingual speech representation learning at scale}.

\bibitem[{Beaufays(2015)}]{Beaufays_2015}
Françoise Beaufays. 2015.
\newblock \href {https://research.google/blog/the-neural-networks-behind-google-voice-transcription/} {The neural networks behind google voice transcription}.

\bibitem[{Bittar et~al.(2024)Bittar, Dixon, Samragh, Nishu, and Naik}]{10447485}
Alexandre Bittar, Paul Dixon, Mohammad Samragh, Kumari Nishu, and Devang Naik. 2024.
\newblock \href {https://doi.org/10.1109/ICASSP48485.2024.10447485} {Improving vision-inspired keyword spotting using dynamic module skipping in streaming conformer encoder}.
\newblock In \emph{ICASSP 2024 - 2024 IEEE International Conference on Acoustics, Speech and Signal Processing (ICASSP)}, pages 10386--10390.

\bibitem[{Calvo et~al.(2014)Calvo, Giménez, Hurtado~Oliver, Sanchis, and Gomez}]{article}
M.~Calvo, M.~Giménez, Lluís Hurtado~Oliver, E.~Sanchis, and Jon Gomez. 2014.
\newblock \href {https://www.researchgate.net/publication/281875726_ELiRF_at_MediaEval_2014_Query_by_example_search_on_speech_task_QUESST} {Elirf at mediaeval 2014: Query by example search on speech task (quesst)}.
\newblock \emph{CEUR Workshop Proceedings}, 1263.

\bibitem[{Dinkel et~al.(2023)Dinkel, Wang, Yan, Zhang, and Wang}]{10095534}
Heinrich Dinkel, Yongqing Wang, Zhiyong Yan, Junbo Zhang, and Yujun Wang. 2023.
\newblock \href {https://doi.org/10.1109/ICASSP49357.2023.10095534} {Unified keyword spotting and audio tagging on mobile devices with transformers}.
\newblock In \emph{ICASSP 2023 - 2023 IEEE International Conference on Acoustics, Speech and Signal Processing (ICASSP)}, pages 1--5.

\bibitem[{Gao et~al.(2023)Gao, Gu, Caliva, and Liu}]{10095929}
Chenyang Gao, Yue Gu, Francesco Caliva, and Yuzong Liu. 2023.
\newblock \href {https://doi.org/10.1109/ICASSP49357.2023.10095929} {Self-supervised speech representation learning for keyword-spotting with light-weight transformers}.
\newblock In \emph{ICASSP 2023 - 2023 IEEE International Conference on Acoustics, Speech and Signal Processing (ICASSP)}, pages 1--5.

\bibitem[{Iqbal et~al.(2020)Iqbal, Zahid, and Raza}]{iqbalunsupervised}
Hafiz~Rizwan Iqbal, Saad~Bin Zahid, and Agha~Ali Raza. 2020.
\newblock \href {https://api.semanticscholar.org/CorpusID:230516072} {An unsupervised spoken term detection system for urdu}.

\bibitem[{Irtza et~al.(2014)Irtza, Dr, and Hussain}]{irtza2014urdu}
Saad Irtza, Khawer~Rehman Dr, and Sarmad Hussain. 2014.
\newblock \href {https://api.semanticscholar.org/CorpusID:207912598} {Urdu keyword spotting system using hmm}.

\bibitem[{Jamal et~al.(2022)Jamal, Abdul~Rauf, and Majid}]{jamal-etal-2022-exploring}
Sahar Jamal, Sadaf Abdul~Rauf, and Quratulain Majid. 2022.
\newblock \href {https://aclanthology.org/2022.eurali-1.11} {Exploring transfer learning for {U}rdu speech synthesis}.
\newblock In \emph{Proceedings of the Workshop on Resources and Technologies for Indigenous, Endangered and Lesser-resourced Languages in Eurasia within the 13th Language Resources and Evaluation Conference}, pages 70--74, Marseille, France. European Language Resources Association.

\bibitem[{Kundu et~al.(2023)Kundu, Samragh, Cho, Padmanabhan, and Naik}]{10097018}
Arnav Kundu, Mohammad Samragh, Minsik Cho, Priyanka Padmanabhan, and Devang Naik. 2023.
\newblock \href {https://doi.org/10.1109/ICASSP49357.2023.10097018} {Heimdal: Highly efficient method for detection and localization of wake-words}.
\newblock In \emph{ICASSP 2023 - 2023 IEEE International Conference on Acoustics, Speech and Signal Processing (ICASSP)}, pages 1--5.

\bibitem[{Patel et~al.(2018)Patel, DN, Fathima, Shah, C, Kumar, and Iyengar}]{Patel_DN_Fathima_Shah_C_Kumar_Iyengar_2018}
Tanvina Patel, Krishna DN, Noor Fathima, Nisar Shah, Mahima C, Deepak Kumar, and Anuroop Iyengar. 2018.
\newblock \href {https://doi.org/10.21437/interspeech.2018-2133} {Development of large vocabulary speech recognition system with keyword search for manipuri}.
\newblock \emph{Interspeech 2018}.

\bibitem[{Pratap et~al.(2023)Pratap, Tjandra, Shi, Tomasello, Babu, Kundu, Elkahky, Ni, Vyas, Fazel-Zarandi, Baevski, Adi, Zhang, Hsu, Conneau, and Auli}]{pratap2023scaling}
Vineel Pratap, Andros Tjandra, Bowen Shi, Paden Tomasello, Arun Babu, Sayani Kundu, Ali Elkahky, Zhaoheng Ni, Apoorv Vyas, Maryam Fazel-Zarandi, Alexei Baevski, Yossi Adi, Xiaohui Zhang, Wei-Ning Hsu, Alexis Conneau, and Michael Auli. 2023.
\newblock \href {http://arxiv.org/abs/2305.13516} {Scaling speech technology to 1,000+ languages}.

\bibitem[{Wei et~al.(2021)Wei, Gong, Yang, Ye, and Wen}]{Wei_Gong_Yang_Ye_Wen_2021}
Yungen Wei, Zheng Gong, Shunzhi Yang, Kai Ye, and Yamin Wen. 2021.
\newblock \href {https://doi.org/10.1007/s12652-021-03022-1} {Edgecrnn: An edge-computing oriented model of acoustic feature enhancement for keyword spotting}.
\newblock \emph{Journal of Ambient Intelligence and Humanized Computing}, 13(3):1525–1535.

\bibitem[{Yang et~al.(2022)Yang, Kim, Chung, and Chang}]{yang2022personalized}
Seunghan Yang, Byeonggeun Kim, Inseop Chung, and Simyung Chang. 2022.
\newblock \href {http://arxiv.org/abs/2206.13708} {Personalized keyword spotting through multi-task learning}.

\end{thebibliography}

\end{document}